\documentclass[prl,reprint,nofootinbib,preprintnumbers,twocolumn,amsmath,amssymb]{revtex4}
\usepackage{graphicx,tikz,braket,booktabs}

\usepackage{appendix}
\newcommand{\beq}{\begin{equation}}
\newcommand{\eeq}{\end{equation}}
\newcommand{\bea}{\begin{eqnarray}}
\newcommand{\eea}{\end{eqnarray}}
\newcommand{\nn}{\nonumber}

\newcommand{\Tr}{\mathop{\rm Tr}}

\def\cN{\mathcal{N}}

\def\det{{\rm det}}
\def\tr{{\rm tr}}
\def\Tr{{\rm Tr}}

\newcommand{\be}{\begin{equation}}
\newcommand{\ee}{\end{equation}}

\newcommand{\hs}[1]{\hspace{#1 mm}}

\def\cC{{\cal C}}
\def\cD{{\cal D}}

\def\cH{{\cal H}}

\def\cN{{\cal N}}

\def\cH{{\cal H}}

\def\cO{{\cal O}}

\def\cN{{\cal N}}
\def\cU{{\cal U}}
\def\cV{{\cal V}}

\def\tr{\mathrm{tr}}
\def\Tr{\mathrm{Tr}}

\newcommand{\br}{\bar R}
\newcommand{\bR}{\bar R}

\begin{document}

\preprint{ACFI-T23-xx}

\title{\bf Physical running of couplings in quadratic gravity }

\medskip\

\medskip\

\author{Diego Buccio${}^1$}
\email{dbuccio@sissa.it}
\author{John F. Donoghue${}^{2}$}
\email{donoghue@physics.umass.edu}
\author{Gabriel Menezes${}^{3}$~\footnote{On leave of absence from Departamento de F\'{i}sica, Universidade Federal Rural do Rio de Janeiro.}}
\email{gabriel.menezes10@unesp.br}
\author{Roberto Percacci${}^1$}
\email{percacci@sissa.it}
\affiliation{
${}^1$International School for Advanced Studies,
Via Bonomea 265, 34134 Trieste, Italy
and INFN, Sezione di Trieste, Trieste, Italy\\
${}^2$Department of Physics,
University of Massachusetts,
Amherst, MA  01003, USA\\
${}^3$Instituto de F\'isica Te\'orica, Universidade Estadual Paulista,
Rua Dr.~Bento Teobaldo Ferraz, 271 - Bloco II, 01140-070 S\~ao Paulo, S\~ao Paulo, Brazil}

\begin{abstract}
We argue that the well-known beta functions of quadratic gravity
do not correspond to the physical dependence of scattering amplitudes
on external momenta, and derive the correct physical beta functions.
Asymptotic freedom turns out to be compatible with the absence of tachyons.
\end{abstract}

\maketitle


Quadratic gravity is an extension of Einstein's theory whose action
contains terms quadratic in curvature.
In signature $-+++$ it reads
\be
S=\int d^4x\sqrt{|g|}
\left[\frac{m_P}{2}^2(R-2\Lambda)
-\frac{1}{2\lambda}C^2
-\frac{1}{\xi}R^2
\right]\ ,
\ee
where $m_P=\sqrt{8\pi G}$ is the Planck mass,
$\Lambda$ is the cosmological constant,
$C_{\mu\nu\rho\sigma}$ is the Weyl tensor.
We will not consider the Euler (Gauss-Bonnet) term here.
This theory is renormalizable \cite{Stelle1}, {and is a potential candidate for a full theory of quantum gravity.}
In addition to the massless graviton it propagates
a massive spin-2 particle that is a ghost and if $\lambda<0$ it is a tachyon\footnote{{Ghosts are particles whose propagator is the negative of the usual one, 
while tachyons are particles with a pole at spacelike momenta.}}.
It also has a massive spin-0 particle that is a tachyon for $\xi>0$.
In spite of these apparent pathologies, it has attracted renewed interest recently
\cite{Salvio:2014soa,Alvarez-Gaume:2015rwa,Holdom:2015kbf,Anselmi:2018ibi,Salvio:2018crh,Donoghue:2021cza,Buoninfante:2023ryt}.
{In these studies, it is suggested that it may be possible that the ghost state is acceptable, although tachyonic states are considered fatal.}

The first attempt to compute  beta functions {for this theory} was made 
by Julve and Tonin in \cite{julve},
but that work missed the contribution of the Nakanishi-Lautrup ghosts.
This was corrected in \cite{ft1} and then,
with some further corrections, in \cite{Avramidi:1985ki}.
The final result is
\bea
\beta_\lambda&=&-\frac{1}{(4\pi)^2}\frac{133}{10}\lambda^2\ ,
\label{abl}
\\
\beta_\xi&=&-\frac{1}{(4\pi)^2}\frac{5(72\lambda^2-36\lambda\xi+\xi^2)}{36}\ ,
\label{abx}
\eea
Since then, these beta functions have been confirmed in several calculations
using different techniques
\cite{BS2,CP,niedermaier,sgrz,Narain:2011gs}. {With these beta functions, full asymptotic freedom can only be
obtained for the case of a tachyonic coupling $\xi>0$.} {The goal of our paper is to recompute these beta functions as appropriate for physical amplitudes and show that in fact full asymptotic freedom can be obtained without tachyons.}

The beta functions (\ref{abl},\ref{abx}) give the dependence of the renormalized
$\lambda$ and $\xi$ on the renormalization scale $\mu$.
We call this the {$\mu${\it-running}.
However, what one is really interested in is the dependence of the
running couplings on external momenta,
that we call {\it physical running}.
\footnote{
These and other definitions of running have been discussed in a simple model
of a higher-derivative shift-invariant scalar theory,
where the full form of the scattering amplitude is accessible \cite{Buccio:2023lzo},
see also \cite{Buccio:2022egr,Tseytlin:2022flu,Holdom:2023usn}.} {Physical scattering amplitudes are independent of $\mu$ after 
renormalization, but do depend on the momenta. In particular, the running of $\lambda$ with momenta enters the spin-two component of the graviton propagator and that of $\xi$ influences the spin-zero propagator.}
{Note that there is no way to define a physical running for the coefficient
of the Euler term, since it does not affect the scattering amplitudes in four dimensions.}

In problems characterized by a single momentum scale $p$,
e.g. the total center of mass energy $p=\sqrt s$,
the $p$-dependence is usually the same as the $\mu$-dependence,
because for dimensional reasons they occur as $\log(p/\mu)$.
In the presence of a non-negligible mass scale $m$,
the amplitude generally contains, 
in addition to terms of the form $\log(p/\mu)$, 
also terms of the form $\log(m/\mu)$ and in this way
the $p$-dependence is no longer correctly reflected
by the $\mu$-dependence.
One clear source of such spurious $\mu$-dependence are tadpoles,
Feynman diagrams that by construction do not depend on the external momenta.
In such cases, the $\mu$-running is not the same as physical running.

{In most familiar quantum field theories such as the Standard Model this is not a problem
as one can use mass independent renormalization schemes. 
However, we claim that it is not always correct in higher derivative theories.}
Two of us have indeed found that in higher derivative sigma models 
the {scale dependent beta functions calculated with a ultraviolet cutoff}
\cite{Hasenfratz:1988rf}
or those obtained from the dependence on an infrared cutoff \cite{Percacci:2009fh}
are indeed contaminated by tadpoles, and hence not physical \cite{Donoghue:2023yjt}.
In the present letter we claim that the same is true
in quadratic gravity, and we compute the physical beta functions.

Calculations of the beta functions so far have been based on the background field
method, expanding $ g_{\mu\nu}=\bar g_{\mu\nu}+h_{\mu\nu} $ around a general background $\bar g$. In the following a bar always indicates a quantity calculated from the background metric.
Almost all calculations used the heat kernel, which is very convenient
because it preserves manifest covariance at all stages. {These are standard techniques and there are many textbooks and reviews approaching the subject, see for instance~\cite{Birrell:1982ix,Barvinsky:1985an,Parker:2009uva,Buchbinder:1992rb,Percacci:2017fkn,Donoghue:2022wrw}.
}

The first step is always the linearization of the action and the choice of a suitable gauge-fixing term,
leading to an action
\be
S^{(2)}=\int d^4x\sqrt{|\bar g|}
h_{\alpha\beta}\cH^{\alpha\beta,\gamma\delta}h_{\gamma\delta}\ .
\ee
One can choose the gauge such that the operator governing
the propagation of gravitons has the form (suppressing the indices),
\be
\cH=\bar\Box^2\mathbb{K}+\mathbb{J}^{\mu\nu}\bar\nabla_\mu\bar\nabla_\nu
+\mathbb{L}^\mu\bar\nabla_\mu+\mathbb{W}
\ee
and $\mathbb{K}$, $\mathbb{J}$, $\mathbb{L}$, $\mathbb{W}$
are matrices in the space of symmetric tensors, depending on $\bR$
and its covariant derivatives.
In particular
\be
\mathbb{K}=\frac{1}{4\lambda}\mathbb{P}_{\textrm{tl}}
+\frac{9}{4(3\xi-2\lambda)}\mathbb{P}_{\textrm{tr}}
\ee
where $\mathbb{P}_{\textrm{tr}}^{\alpha\beta,\gamma\delta}
=\frac14\bar g^{\alpha\beta}\bar g^{\gamma\delta}$
is the projector on the trace part and
$\mathbb{P}_{\textrm{tl}}=\mathbb{I}-\mathbb{P}_{tr}$
the projector on the traceless part.
In flat space, $\mathbb{K}$ can be viewed as a tensorial
wave function renormalization constant that gives different
weights to the spin-2 and spin-0 components of $h$.
As usual it is convenient to canonically normalize the fields
by redefining $h\to \sqrt{\mathbb{K}^{-1}}h$,
so that the action can be rewritten as 
\be
S^{(2)}=\int d^4x\sqrt{|\bar g|}
h_{\alpha\beta}\cO^{\alpha\beta,\gamma\delta}h_{\gamma\delta}\ ,
\label{Slin}
\ee
where, suppressing again the indices,
\be
\cO=\bar\Box^2\mathbb{I}+\mathbb{V}^{\mu\nu}\bar\nabla_\mu\bar\nabla_\nu
+\mathbb{N}^\mu\bar\nabla_\mu+\mathbb{U}\ ,
\label{operator}
\ee
and $\mathbb{V}=\sqrt{\mathbb{K}^{-1}}\mathbb{J}\sqrt{\mathbb{K}^{-1}}$ etc.
Now $\mathbb{V}$ contains terms proportional to $\bR$ and $m_P^2$,
$\mathbb{N}$ contains terms proportional to $\bar\nabla\bR$,
whereas $\mathbb{U}$ contains terms proportional to $\bR^2$, $\bar\nabla^2\bR$, $m_P^2 \bR$ and $m_P^2\Lambda$.
The logarithmic divergences, or equivalently the $1/\epsilon$ poles
in dimensional regularization, are proportional to
the heat kernel coefficient \cite{Barvinsky:1985an}
\bea
&&\hs{-4}
\frac{1}{32\pi^2}
\int d^4x\,\mathrm{tr}\Bigl[
\frac{\mathbb{I}}{90} \left(\br_{\rho\lambda\sigma\tau}^2 
-\br_{\rho\lambda}^2
+\frac52\br^2 \right)
+\frac{1}{6}\mathbb{R}_{\rho\lambda}\mathbb{R}^{\rho\lambda}
\nonumber\\
&& \hs{-4} 
 - \frac{\br_{\rho\lambda}\mathbb{V}^{\rho\lambda}
- \tfrac12\br \mathbb{V}^\rho{}_\rho }{6}
+ \frac{\mathbb{V}_{\rho\lambda} \mathbb{V}^{\rho\lambda}
+\tfrac12\mathbb{V}^\rho{}_\rho \mathbb{V}^\lambda{}_\lambda
}{24} - \mathbb{U}\Bigr]\ ,
\label{grav}
\eea
where $\mathbb{R}_{\rho\lambda}=[\nabla_\rho,\nabla_\lambda]$
acting on symmetric tensors.

The operator (\ref{operator}) is of the same form as the one
acting on the scalars in the higher derivative sigma models, 
and so are the divergences (\ref{grav}) (except for the terms $\tr \bR\mathbb{V}$).
We can then use the same arguments of \cite{Donoghue:2023yjt}.
The terms in the first line of (\ref{grav}) are the ones that we would
get for $\cO=\bar\Box^2$.
Using the formula $\Tr\log\Box^2=2\Tr\log\Box$
one can conclude that none of those terms could be a tadpole,
because with a standard $\/p^2$ propagator
a diagram must have at least two propagators to be logarithmically divergent.

On the other hand consider {the term in Eq. \ref{grav} which is linear in the $\mathbb{U} $ interaction, i.e.  $\tr\mathbb{U}$. As this only involves one vertex, it is clear that the loop
diagram involved must be a tadpole diagram as seen in the second diagram of Fig. 1.}
Some more detailed arguments lead to the conclusion that also
some of the $\tr\bR\mathbb{V}$ divergences are due to tadpoles.
This is enough to conclude that the standard beta functions (\ref{abl},\ref{abx})
cannot be the physical ones.

We thus wish to evaluate the physical beta functions.
In order to use flat space Feynman diagrams,
we go back to the original Julve-Tonin approach
and assume that the background is itself a small deformation of flat space
$\bar g_{\mu\nu}=\eta_{\mu\nu}+f_{\mu\nu}\ .$
Expanding around flat space, the action (\ref{Slin})
gives rise to an operator of the form
\bea
\cO &\equiv& \boxdot^2\mathbb{I}
+\cD^{\mu\nu\rho\sigma}\partial_\mu\partial_\nu\partial_\rho\partial_\sigma
+\cC^{\mu\nu\rho}\partial_\mu\partial_\nu\partial_\rho
\nn\\
&+&\cV^{\mu\nu}\partial_\mu\partial_\nu
+\cN^\mu\partial_\mu+\cU\ ,
\label{operatorf}
\eea
where $\boxdot$ is the flat Laplacian, $\cD$ and $\cC$ come from the expansion of $\sqrt{\bar g}\bar\Box^2$
and $\cV$, $\cN$ and $\cU$ are equal to $\mathbb{V}$, $\mathbb{N}$ and $\mathbb{U}$
plus terms coming from the expansion of $\sqrt{\bar g}\bar\Box^2$.
Each of these terms is an infinite series in $f$. 
{Recall that the functional trace of the logarithm of an operator can be approximated by
\bea
\hspace{-5mm}
\mathrm{tr}\log \cO\!\! &=&\!\!
\mathrm{tr}\log\left(\boxdot^2+A \right)
\nonumber\\
\!\!&\approx&\!\! \mathrm{tr}\left[2\log\boxdot+A \frac{1}{\boxdot^2}
-\frac12 A \frac{1}{\boxdot^2} A \frac{1}{\boxdot^2}+ \cdots \right] .
\label{logexp}
\eea
In the above expansion, $A$ generically represents the remaining contributions to $\cO$ appearing in Eq.~(\ref{operatorf}), and again we are suppressing Lorentz indices for brevity. Furthermore, the first term in the perturbative expansion of Eq.~(\ref{logexp}) corresponds to tadpole integrals, while the second term can be evaluated as a bubble Feynman diagram. Our discussion here concerns how to compute terms proportional to $\log p^2$.}

The physical running of $\lambda$ and $\xi$ comes from terms
\be
b_\lambda \bar C^{\mu\nu\rho\sigma}\log\bar\Box \bar C_{\mu\nu\rho\sigma}
+b_\xi \bR\log\bar\Box \bR
\label{ea}
\ee
in the effective action, and the beta functions are
$$
\beta_\lambda=-4b_\lambda\lambda^2\ ,\ \ 
\beta_\xi=-2b_\xi\xi^2\ .
$$
In flat space contributions to the coefficients $b_\lambda$ and $b_\xi$ can be read from
the two point function of the background fluctuation $f$,
which is represented graphically by the diagrams in Figure \ref{fig:feynman}. 

\usetikzlibrary {arrows.meta}
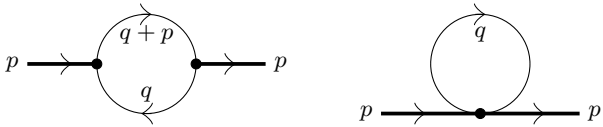
\begin{figure}[ht]
\begin{center}
\begin{tikzpicture}
[>=stealth,scale=0.66,baseline=-0.1cm]
\draw (-1,0) arc (180:0:1cm);
\draw (-1,0) arc (180:360:1cm);
\draw [line width=0.5mm ] (-1,0) node[anchor=west] {} -- (-2.4,0);
\draw [line width=0.5mm ] (2.4,0) -- (1,0) node[anchor=east] {};
\draw [-{Classical TikZ Rightarrow[length=1.5mm]}] (-1.6,0) -- (-1.5,0);
\draw [-{Classical TikZ Rightarrow[length=1.5mm]}] (1.7,0) -- (1.8,0);
\draw [-{Classical TikZ Rightarrow[length=1.5mm]}] (-0.1,1) -- (0.1,1);
\draw [-{Classical TikZ Rightarrow[length=1.5mm]}] (0.1,-1) -- (-0.1,-1);
\draw[fill] (1,0) circle [radius=0.1];
\draw[fill] (-1,0) circle [radius=1mm];
\node [] at (-2.7,0) {$p$};
\node [] at (2.7,0) {$p$};
\node [] at (0,0.6) {$q+p$};
\node [] at (0,-0.6) {$q$};
\end{tikzpicture}
\qquad
\begin{tikzpicture}
[>=stealth,scale=0.66,baseline=-0.1cm]
\draw (-1,0) arc (180:0:1cm);
\draw (-1,0) arc (180:360:1cm);
\draw [line width=0.5mm ](2,-1) node[anchor=west] {} -- (0,-1);
\draw [line width=0.5mm ](0,-1) -- (-2,-1) node[anchor=east] {};
\draw [-{Classical TikZ Rightarrow[length=1.5mm]}] (-1.2,-1) -- (-1.1,-1);
\draw [-{Classical TikZ Rightarrow[length=1.5mm]}] (1.2,-1) -- (1.3,-1);
\draw [-{Classical TikZ Rightarrow[length=1.5mm]}] (-0.1,1) -- (0.1,1);
\draw[fill] (0,-1) circle [radius=1mm];
\node [] at (-2.3,-1) {$p$};
\node [] at (2.3,-1) {$p$};
\node [] at (0,0.6) {$q$};
\end{tikzpicture}
\caption{Diagrams contributing to the two-point function:
bubbles (left) and tadpoles (right).
The thin line can be the $h$ propagator or one of the ghosts, the thick line is the $f$ propagator, with momentum $p$.
The vertices can come from expanding any one among $\cD$, $\cC$, $\cV$, $\cN$, $\cU$.
}
\label{fig:feynman}
\end{center}
\end{figure}

The two $h$-$h$-$f$ vertices in the bubble diagrams are obtained
by expanding $\cD$, $\cC$, $\cV$, $\cN$ and $\cU$ to first order,
while for the tadpole one has to expand to second order.
Being logarithmically divergent, the tadpole contributes to the $\mu$-running
but not to the $p$-dependence that we are interested in.
Thus the bulk of the calculation consists of working out the
Feynman integrals for each of the 15 possible pairs of vertices in the bubble
and then evaluating the result for the specific form of the operator (\ref{operatorf}). 

The calculation is simplified by neglecting the terms proportional to $m_P$.
This is justified in the UV limit,
as seen explicitly
in the case of the simple shift-invariant scalar model
\cite{Buccio:2023lzo}.
{The calculation of the relevant Feynman integrals becomes straightforward and the results are given in the 
Supplemental Material, where we also present all possible pairs of vertices appearing in the bubble integral.}

In the calculation one sees in detail how it happens that the $\mu$-running differs
from the physical running.
In dimensional regularization the $\log\mu$ terms always appear
together with the $1/\epsilon$ pole, so the $\mu$-running just
traces the log divergences of the theory.
We have checked that putting together all the bubble {\it and} tadpole
diagrams one reconstructs the covariant expression (\ref{ea})
with the coefficients leading to the standard beta functions (\ref{abl},\ref{abx}).
If we just drop the tadpoles, the resulting function of $f$ is not
the linearization of a covariant expression.
Thus, the physical running cannot be obtained from the $\mu$-running
by just dropping the tadpole contribution.
Instead, there are other contributions.

As we have explained earlier, in the presence of a mass,
the $\mu$-dependence does not correctly describe the amplitude.
In our theory the only mass is the Planck mass and one would expect
that in the limit $p \gg m_P$, it becomes negligible.
However, {in this theory with four-derivative propagators the Planck mass also keeps the theory infrared finite.  If we neglect $m_P$ in the limit $p \gg m_P$ limit, one finds infrared divergences.}
This is a new phenomenon that does not occur in standard
two-derivative theories.
There are then two { simple ways to deal with this situation.}
{One is to continue to use dimensional regularization to regulate also the IR divergences {which appear in the $m_P\to 0$ limit.}
In this case {\it all} the logs are again of the form $\log p^2/\mu^2$,
but in addition to the UV logs there are now also IR logs,
that change the beta function.
As we explain more fully below, summing all the $\log p^2$ terms now gives again a covariant
expression, but with a different coefficient.
This is the physical beta function.
One could alternatively reintroduce artificially a small mass $m$ as an IR regulator
{by letting $q^4\to q^4+m^2q^2$.}
\footnote{{If we had kept the Hilbert term, $m^2$ would be either
$\lambda m_P^2$ or $\xi m_P^2$ depending on the spin component of the propagator. However, the physical running with $\log p^2$ is independent
of these masses so that it is most convenient to just use a common mass when using it as an IR regulator. No spurious poles are introduced by this procedure as long as the sign of $m^2$ is chosen to avoid tachyons.}}
The presence of the regulator mass leads to 
terms of the form $\log p^2/m^2$, and we are interested in the $\log p^2 $ effects.}
We have checked that both procedures lead exactly to the same result.
Notice that the small-time expansion of the heat kernel always gives only the
UV divergences.

In our diagrams the IR divergences always appear with powers of
the external momentum $p$ in the denominator and therefore give rise
to apparently nonlocal $1/\Box$ or $1/\Box^2$ terms.
However, since the interactions always involve derivatives,
they are offset by an equal number of powers of $p$ in the numerator.
Due to differential identities such as
\bea
\bar\nabla_\mu\bar\nabla_\nu\bR^{\mu\nu\rho\sigma}
\bar\nabla_\alpha\bar\nabla_\beta\bR^{\alpha\beta}{}_{\rho\sigma}
&=&\bR_{\mu\nu}\bar\Box^2\bR^{\mu\nu}
-\frac14 \bR\bar\Box^2\bR
\nn\\
&+& O(\bR^3)\ ,
\eea
or their linearized versions,
these momenta always appear in the combination $p^2$ and
cancel the inverse powers of $p$.
In this way also the logs of infrared origin {appear as coefficients of} local operators.
However, these operators are not by themselves the linearization of a covariant expression.
It is only when one adds them to the UV logs that they give rise
to a covariant expression as in (\ref{ea}). {Both types of logs are physical and both are needed to maintain general covariance.}

These points can be seen easily by considering the vertex $\mathbb{U}$.
It enters the heat kernel calculation linearly, see Eq. (\ref{grav}),
corresponding to a tadpole. It is only the part of $\mathbb{U}$
quadratic in curvature that contributes to the beta functions (\ref{abl},\ref{abx}) {describing $\mu$ running}.
By contrast in our calculation we need two powers of $\mathbb{U}$
and hence only the part proportional to $\bar\nabla\bar\nabla\bR$
contributes at order $f^2$.
These $\mathbb{U}$-$\mathbb{U}$ bubbles are UV finite
but contain $\log p/m$ contributions coming from the IR region.
These come with a factor $p^4$ in the denominator, from the propagators,
but also $p^4$ in the numerator from the vertices.
Thus they contribute to the terms (\ref{ea}).

Finally we observe that with our choice of gauge,
the Faddeev-Popov ghost operators are of second order in derivatives and none of these
exotic phenomena can happen.
Thus their contribution can be taken from traditional heat kernel calculations.

Putting everything together, our final result is
\bea
\beta_\lambda&=&-\frac{1}{(4\pi)^2}\frac{(1617\lambda-20\xi)\lambda}{90}\ ,
\label{bdmpl}
\\
\beta_\xi&=&-\frac{1}{(4\pi)^2}\frac{\xi^2-36\lambda\xi-2520\lambda^2}{36}\ ,
\label{bdmpx}
\eea
The flowlines around the free fixed point $\lambda=\xi=0$ are shown
in Figure \ref{fig:flow}.

\begin{figure}[ht]
\begin{center}
\includegraphics[scale=0.5]{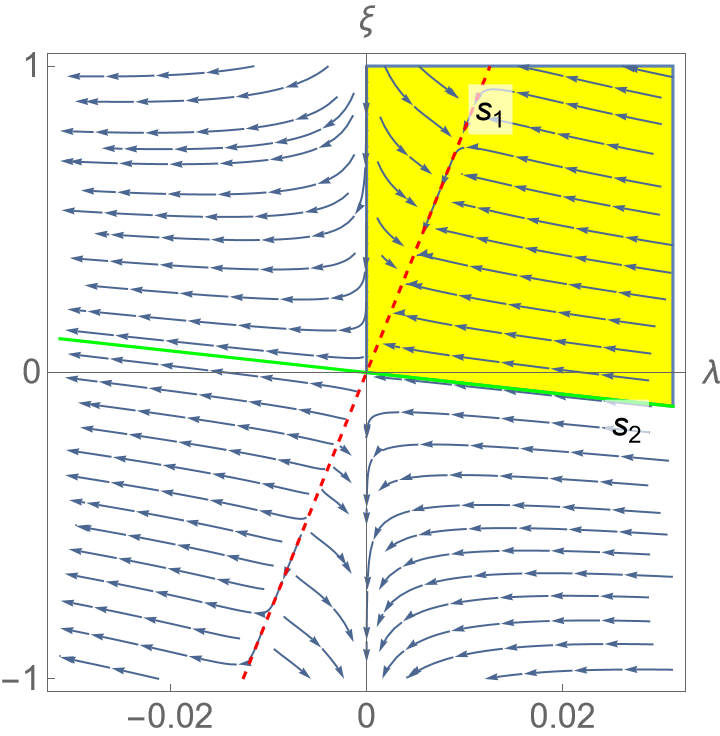}
\caption{Flowlines of the beta functions (\ref{bdmpl},\ref{bdmpx}).
The red dashed line corresponds to (\ref{s1}) and the green line to (\ref{s2}).
Initial points in the shaded area are asymptotically free. In the two left quadrants
the massive spin 2 state is a tachyon, in the two upper quadrants
the massive spin 0 state is a tachyon.}
\label{fig:flow}
\end{center}
\end{figure}

There are three separatrices, along which the motion is purely radial.
The line $\lambda=0$ is UV repulsive for $\xi>0$ and UV attractive for $\xi<0$;
the line $s_1$ is defined by
\be
\xi=\frac{569+\sqrt {386761}}{15}\lambda\approx 79.4\lambda
\label{s1}
\ee
and is attractive for $\lambda>0$ and repulsive for $\lambda<0$,
and the line $s_2$ is defined by
\be
\xi=\frac{569-\sqrt {386761}}{15}\lambda\approx -3.53\lambda
\label{s2}
\ee
and is repulsive for $\lambda>0$ and attractive for $\lambda<0$.
Thus the region that is attracted towards the free fixed point
is the upper right quadrant, plus a triangular slice of the lower right quadrant
that lies above the separatrix $s_2$.

Recall that absence of tachyons requires $\lambda>0$ and $\xi<0$.
There is a unique trajectory that is asymptotically free
and lies entirely in the tachyon-free area, and that is the separatrix $s_2$. This behavior is to be contrasted with 
the flow {related to  $\mu$ running in Eqns.} (\ref{abl},\ref{abx}),
for which the analog of the separatrix $s_2$ is the only asymptotically free trajectory,
but with a positive slope,
with the result that it lies entirely in the tachyonic region. {The physical running couplings
allow asymptotic freedom without tachyons. Moreover there may
be an additional possibility. {The pole in the spin-zero propagator appears at $p^2=-\xi(p^2) m_P^2$, so that  one can also have asymptotically free trajectories where the coupling
$\xi (p^2)$ is negative at the pole in order to avoid a tachyonic state, but has a positive sign at higher momenta.   One can see such trajectories 
above $s_2$ which have $\xi>0$ in the far UV
but eventually cross into $\xi<0$ when one goes towards the IR.}

{In summary, {a key result of our work is that asymptotic freedom can be obtained in quadratic gravity without tachyons. This is because} the physical running with the momenta is the appropriate running coupling to be used in amplitudes. We have noted the difference between this running and that which just follows $\log \mu$ when applied to theories with  higher derivatives, {and have developed new methods to calculate the physical running in the gravitational theory.  The possiblility of tachyonless asymptotic freedom makes quadratic gravity more plausible as a complete renormalizeable theory of quantum gravity.}}

{\it Acknowledgements} --  We thank Andrei Barvinsky, Arkady Tseytlin and Alexander Monin for useful comments or discussions. JFD acknowledges partial support from the U.S. National Science Foundation under grant NSF-PHY-21-12800. GM acknowledges partial support from Conselho Nacional de Desenvolvimento Cient\'ifico e Tecnol\'ogico - CNPq under grant 317548/2021-2, Funda\c{c}\~ao Carlos Chagas Filho de Amparo \`a Pesquisa do Estado de S\~ao Paulo - FAPESP under grant 2023/06508-8 and Funda\c{c}\~ao Carlos Chagas Filho de Amparo \`a Pesquisa do Estado do Rio de Janeiro - FAPERJ under grant E-26/201.142/2022.

\onecolumngrid
\appendix

\section{Appendix -- High energy limit of $\log(p^2)$ coefficients in bubble integrals}

In this Appendix we review some basic notions of the background-field method. We also collect all the 15 possible pairs of vertices and all the associated $\log p^2$ terms emerging from relevant Feynman integrals discussed in the main text.

Let us begin our discussion by reviewing the one-loop effective action associated with a massless scalar field propagating in a curved background. In the background-field method, the action $S[\phi]$ is written in term of a classical background field $\bar\phi$ and a quantum fluctuation $\varphi$ as $S[\bar\phi, \varphi]$, where $\phi=\bar\phi+\varphi$. In this representation, the effective action will be 
\be
e^{-\Gamma[\bar\phi]}=\int D\varphi \, e^{-S[\bar\phi, \varphi]}
\ee
Near to a configuration of the classical field $\bar\phi_0$ which is a solution of equations of motion, and hence a minimum of the classical action, the action can be approximated by the action computed in its minimum and a quantum part, quadratic in quantum fluctuations, equal to the Hessian of the action computed in the minimum:
\be
S[\bar\phi_0,\varphi]\approx S[\bar\phi_0]+\frac12 \varphi\frac{\delta S[\phi]}{\delta \phi \delta \phi}|_{\phi=\bar\phi_0}\varphi
\ee
So the effective action is given by
\be
e^{-\Gamma[\bar\phi]}=e^{-S[\bar\phi]}\int D\varphi e^{-\frac12 \varphi \cO\varphi}
\ee
were we have defined 
$$
\cO=\frac{\delta S[\phi]}{\delta \phi \delta \phi}|_{\bar\phi=\bar\phi_0}.
$$ 
The functional integral is now just a Gaussian integral in $\varphi$ and is equal to $(\det\cO)^{-\frac12}$. We can write the one-loop corrected effective action as
\be
\Gamma[\bar\phi]=S[\bar\phi]+\frac12 \log\det\cO
\ee
Moreover, $\log\det\cO=\mathrm{tr}\log\cO$, so $\Delta_1\Gamma=\frac12\mathrm{tr}\log\cO$.

\begin{table}[ht]
\begin{tabular}{lcc}
\hline
vertices & numerator & $\log(p^2)$ term \\ \hline
    $UU$     &     $-\frac12UU$      &   $ -\frac{UU}{16 \pi^2 p^4} $     \\ \hline
    $UN$     &     $-\frac i2 U N^\mu q_\mu$      & $i\frac{U N^\mu p_\mu}{32 \pi ^2 p^4}$         \\ \hline
    $UV$     &    $\frac12U V^{\mu\nu} q_\mu q_\nu$      &   $\frac{ U V^{\mu\nu} p_\mu p_\nu}{32 \pi ^2 p^4}$       \\ \hline
    $UC$     &    $ \frac i2 U C^{\mu\nu\rho}q_\mu q_\nu q_\rho $     &   $ -i\frac{  U C^{\mu\nu\rho}p_\mu p_\nu p_\rho}{32 \pi ^2 p^4}  $    \\ \hline
    $UD$     &    $-\frac12U D^{\mu\nu\rho\sigma} q_\mu q_\nu q_\rho q_\sigma $       &    $-\frac{1}{32 \pi ^2
   }\left(\frac{1}{p^4}U D^{\mu\nu\rho\sigma} p_\mu p_\nu p_\rho p_\sigma-\frac18U D^{\mu\ \nu}_{\ \mu\ \nu}\right)$      \\ \hline
    $NN$     &    $-\frac12 N^\mu N^\nu q_\mu (q+p)_\nu$       &    $0$     \\ \hline
    $NV$     &    $-\frac i2 N^\mu V^{\nu\rho} (p+q)_\mu q_\nu q_\rho$       &    $0$      \\ \hline
    $NC$     &     $\frac12 N^\sigma C^{\mu\nu\rho} q_\mu q_\nu q_\rho (q+p)_\sigma$     &     $-\frac{ N^\mu C_{\mu}{}^{\nu}{}_{\nu} }{256 \pi ^2}$     \\ \hline
    $ND$     &     $\frac i2 N^\lambda D^{\mu\nu\rho\sigma}q_\mu q_\nu q_\rho q_\sigma(q+p)_\lambda
   $      &    $\frac{i}{128 \pi ^2
   }\left( p_\lambda N_\mu  D^{\mu\lambda\nu}{}_{\nu}-\frac{ p_\lambda N^\lambda  D^{\mu\ \nu}_{\ \mu\ \nu}}{4}\right)$      \\ \hline
    $VV$     &    $-\frac12 V^{\mu\nu} V^{\rho\sigma} q_\mu q_\nu  (q+p)_\rho(q+p)_\sigma$       &     $\frac{1}{384 \pi ^2}\left(V^{\mu\nu}V_{\mu\nu}+\frac{ V^\mu_{\ \mu} V^\nu_{\ \nu}}{2 }\right)$     \\ \hline
    $VC$     &      $-\frac i2 V^{\sigma\lambda} C^{\mu\nu\rho} q_\mu q_\nu q_\rho (q+p)_\sigma (q+p)_\lambda$     &  \begin{tabular}[c]{@{}l@{}}$-\frac{i}{256 \pi ^2
   }\Big(- p_\mu
   V^{\mu}_{\ \nu} C^{\nu\rho}_{\ \ \rho}+\frac{ p_\mu V^\nu_{\ \nu} C^{\mu\rho}_{\ \ \rho}}{2}+ p_\mu
   V_{\nu\rho} C^{\mu\nu\rho}\Big)$\end{tabular}        \\ \hline
    $VD$     &   \begin{tabular}[c]{@{}l@{}}  $\frac12 V^{\lambda\delta} D^{\mu\nu\rho\sigma}q_\mu q_\nu q_\rho q_\sigma\times$\\$(q+p)_\lambda(q+p)_\delta $ \end{tabular}     &   \begin{tabular}[c]{@{}l@{}}$\frac{1}{160 \pi ^2
   }\Big(p_\mu p_\nu V^\mu_{\ \rho} D^{\nu\rho\sigma}_{\ \ \ \sigma}-\frac{3  V^{\mu\nu} p_\mu p_\nu  D^{\rho\ \sigma}_{\ \rho\ \sigma}}{16}-\frac{3 p_\mu p_\nu V^\rho_{\ \rho} D^{\mu\nu\sigma}_{\ \ \ \sigma}}{8}$\\ $-\frac{3 p_\mu p_\nu V_{\rho\sigma} D^{\mu\nu\rho\sigma}}{4}+\frac{p^2 V^\mu_{\ \mu} D^{\nu\ \rho}_{\ \nu\ \rho}}{16 }+\frac{p^2 V_{\mu\nu}D^{\mu\nu\rho}_{\ \ \ \rho}}{4}$\Big) \end{tabular}       \\ \hline
    $CC$     & \begin{tabular}[c]{@{}l@{}}   $-\frac12 C^{\mu\nu\rho} C^{\sigma\lambda\delta} q_\mu q_\nu q_\rho \times$\\$(q+p)_\sigma(q+p)_\lambda(q+p)_\delta$ \end{tabular}       &  \begin{tabular}[c]{@{}l@{}}$-\frac{1}{640 \pi ^2
   }\Big(\frac{ C^{\mu\nu\rho} C_{\mu\nu\rho}p^2}{2}+\frac{ 3C^{\mu\ \nu}_{\ \nu\ } C_{\mu\ \rho}^{\ \rho\ }p^2}{4}+$\\ $3  p_\mu p_\nu C^{\mu\rho\sigma}C^\nu_{\ \rho\sigma}+  \frac{3
    p_\mu p_\nu C^{\mu\rho}_{\ \ \rho}C^{\nu\sigma}_{\ \ \sigma}}{2}-\frac{9
    p_\mu p_\rho C^{\mu\rho}_{\ \ \nu}C^{\nu\sigma}_{\ \ \sigma}}{2}$\Big)\end{tabular}      \\ \hline
    $CD$     &  \begin{tabular}[c]{@{}l@{}}  $-\frac i2  C^{\lambda\delta\alpha} D^{\mu\nu\rho\sigma}q_\mu q_\nu q_\rho q_\sigma\times$\\$(q+p)_\lambda(q+p)_\delta(q+p)_\alpha$ \end{tabular}       & \begin{tabular}[c]{@{}l@{}}$-\frac{i}{320 \pi ^2
   }\Big(-\frac{3 C^{\rho\sigma\delta} D^{\mu\nu}_{\ \ \sigma\delta}p_\mu p_\nu p_\rho}{2}-\frac{3  C^{\rho\sigma}_{\ \ \sigma} D^{\mu\nu\delta}_{\ \ \ \delta}p_\mu p_\nu p_\rho}{4}+$ \\ $
   C^{\sigma}_{\ \sigma\delta} D^{\mu\nu\rho\delta}p_\mu p_\nu p_\rho+\frac{3 C^{\mu\nu}_{\ \ \sigma} D^{\rho\sigma\delta}_{\ \ \ \delta}p_\mu p_\nu p_\rho}{2}-\frac{ C^{\mu\nu\rho} D^{\sigma\ \delta}_{\ \sigma \ \delta}p_\mu p_\nu p_\rho}{4}+$ \\ $ \frac{3 p^2 p_\mu
    C^{\mu\nu\rho} D^{\ \ \delta}_{\nu\rho \ \delta}}{4}+\frac{3 p^2 p_\mu
    C^{\mu\nu}_{\ \ \nu} D^{\rho \ \sigma}_{\ \rho \ \sigma}}{16}-\frac{3 p^2 p_\mu
    C^{\nu\ \rho}_{\ \nu} D^{\mu\ \delta}_{\ \rho \ \delta}}{4}-$ \\ $\frac{ p^2 p_\mu
    C_{\nu\rho\sigma} D^{\mu\nu\rho\sigma}}{2}\Big)$\end{tabular}        \\ \hline
    $DD$     &   \begin{tabular}[c]{@{}l@{}}  $-\frac12   D^{\mu\nu\rho\sigma}D^{\lambda\delta\alpha\beta}q_\mu q_\nu q_\rho q_\sigma\times$\\$(q+p)_\lambda(q+p)_\delta(q+p)_\alpha(q+p)_\beta$    \end{tabular}   &     \begin{tabular}[c]{@{}l@{}} 
    $-\frac{1}{140 \pi ^2
   }\Big(-\frac{9  D^{\mu\nu}_{\ \ \alpha\beta}D^{\rho\sigma\alpha\beta}p_\mu p_\nu p_\rho p_\sigma}{16 }-\frac{9  D^{\mu\nu\lambda}_{\ \ \ \lambda}D^{\rho\sigma\delta}_{\ \ \ \delta}p_\mu p_\nu p_\rho p_\sigma}{32 }+$ \\ $ D^{\mu\nu\rho\delta}D^{\sigma\ \lambda}_{\ \delta\ \lambda}p_\mu p_\nu p_\rho p_\sigma-\frac{5 D^{\mu\nu\rho\sigma}D^{\lambda\ \delta}_{\ \lambda \ \delta}p_\mu p_\nu p_\rho p_\sigma}{32}-$ \\ $\frac{3 p^2 p_\mu p_\nu D^{\mu\rho\sigma\lambda}   D^{\nu}_{\ \rho\sigma\lambda}}{8}-\frac{9 p^2 p_\mu p_\nu D^{\mu\rho\ \lambda}_{\ \ \rho}   D^{\nu\ \sigma}_{\ \lambda\ \sigma}}{16}+$ \\ $\frac{3 p^2 p_\mu p_\nu D^{\mu\nu\rho\sigma}   D^{\ \ \lambda}_{\rho\sigma\ \lambda}}{4 }+\frac{3 p^2 p_\mu p_\nu D^{\mu\nu\rho}_{\ \ \ \rho}  D^{\sigma\ \lambda}_{\ \sigma\ \lambda}}{16 }-$ \\ $\frac{3 p^4 D^{\mu\nu\rho\sigma}D_{\mu\nu\rho\sigma}}{128 }-\frac{9 p^4 D^{\mu\ \rho\sigma}_{\ \mu}D^{\ \ \nu}_{\rho\sigma\ \nu}}{128}-\frac{9 p^4 D^{\mu\ \nu}_{\ \mu\ \nu}D^{\rho\ \sigma}_{\ \rho\ \sigma}}{1024 }\Big)$ 
   \end{tabular} \\
   \hline
\end{tabular}
\caption{Scalar bubble loops.}
\label{tab2}
\end{table}

Let's consider a quadratic operator with structure
\be
\cO=\boxdot^2+D^{\mu\nu\rho\sigma}\partial_\mu\partial_\nu\partial_\rho\partial_\sigma+C^{\mu\nu\rho}\partial_\mu\partial_\nu\partial_\rho+V^{\mu\nu}\partial_\mu\partial_\nu+N^\mu\partial_\mu+U \label{operator1}
\ee
where all terms have mass dimension 4. After a Fourier transform, one can build 15 different bubble integrals with the interaction vertices $D, C, V, N$ and $U$, which differ from each other only by the numerator in the momentum integral and can be computed using standard Feynman integrals. 
All bubbles are symmetric under the exchange of vertices, as can be easily verified using the variable redefinition $q\to-q-p$ and considering that $U$, $V$ and $D$ are invariant when moved from the left to the right of the diagram, while $N$ and $C$ go to $-N$ and $-C$, since the external momentum $p$ is ingoing on the left and outgoing on the right.

The contribution from bubble diagrams to the one-loop effective action is 
\be
UU+NN+VV+CC+DD+2(UV+UN+UC+UD+NV+NC+ND+VC+VD+CD).
\label{sum0}
\ee
If $\cO$ is a self-adjoint operator, that means $\braket{x|\cO y}=\braket{\cO x|y}$. 
The symmetry of $\cO$ permits us to take it as the inverse propagator and compute only the particular contraction between vertices dictated by the expansion of the functional trace of the logarithm of the operator $\cO$
\bea
&&\mathrm{tr}\log\left(\cO_{sym} \right)
\approx\mathrm{tr}\log\left(\boxdot^2\right)+\mathrm{tr}\left[(D_{\rho\lambda\alpha\beta} \partial^\rho \partial^\lambda \partial^\alpha \partial^\beta+ \cdots)\frac{1}{\boxdot^2}-\right.
\nonumber\\
&&\left.\frac12(D_{\rho\lambda\alpha\beta} \partial^\rho \partial^\lambda \partial^\alpha \partial^\beta+ \cdots)\frac{1}{\boxdot^2}(D_{\rho\lambda\alpha\beta} \partial^\rho \partial^\lambda \partial^\alpha \partial^\beta+ \cdots)\frac{1}{\boxdot^2}+ \cdots  \right] + \cdots
\eea
where the dots represent the remaining contributions not displayed in the above equation for clarity. In addition, we have indicated by $\cO_{sym}$ the result of computing the contraction just mentioned. We report in table \ref{tab2} the high energy limit of the part proportional to $\log(p^2)$ in all the possible bubble diagrams.

\begin{table}[]
\begin{tabular}{lcc}
\hline
vertices & numerator & $\log(p^2)$ term \\ \hline
    $\cU\cU$     &     $-\frac12\cU_{AB}\cU^{BA}$      &   $   -\frac{\cU^{AB}\cU_{BA}}{16 \pi^2 p^4}$     \\ \hline
    $\cU \cN$     &     $-\frac i2 \cU_{AB} \cN^\mu{}^{BA} q_\mu$      & $i\frac{\cU^{AB} \cN^\mu_{BA} p_\mu}{32 \pi ^2 p^4}$         \\ \hline
    $\cU \cV$     &    $\frac12\cU_{AB} \cV^{\mu\nu}{}^{BA} q_\mu q_\nu$      &   $\frac{ \cU^{AB} \cV^{\mu\nu}_{BA} p_\mu p_\nu}{32 \pi ^2 p^4}$       \\ \hline
    $\cU \cC$     &    $ \frac i2 \cU_{AB} \cC^
{\mu\nu\rho}{}^{BA}q_\mu q_\nu q_\rho$     &   $ -i\frac{  \cU^{AB} \cC^{\mu\nu\rho}_{BA}p_\mu p_\nu p_\rho}{32\pi ^2 p^4} $    \\ \hline
    $\cU \cD$     &    $-\frac12\cU_{AB} \cD^{\mu\nu\rho\sigma}{}^{BA} q_\mu q_\nu q_\rho q_\sigma $       &    $-\frac{1}{32 \pi ^2
   }\Big(\frac{\cU^{AB} \cD^{\mu\nu\rho\sigma}_{BA} p_\mu p_\nu p_\rho p_\sigma}{p^4}-\frac{\cU^{AB} \cD^{\mu\ \nu}_{\ \mu\ \nu}{}_{BA}}{8}\Big)$     \\ \hline
    $\cN\cN$     &    $-\frac12 \cN^\mu_{AB} \cN^\nu{}^{BA} q_\mu (q+p)_\nu$       &    $0$     \\ \hline
    $\cN\cV$     &    $-\frac i2 \cN^\mu_{AB} \cV^{\nu\rho}{}^{BA} q_\nu q_\rho(p+q)_\mu $       &    $0$      \\ \hline
    $\cN\cC$     &  $\frac12 \cN^\sigma_{AB} \cC^{\mu\nu\rho}{}^{BA} q_\mu q_\nu q_\rho (q+p)_\sigma$    &   $-\frac{\cN^\sigma{}^{AB} \cC^{\ \mu}_{\sigma\ \mu}{}_{BA}}{256 \pi ^2 }$   \\ \hline
    $\cN \cD$     &    $\frac i2 \cN^\lambda_{AB} \cD^{\mu\nu\rho\sigma}{}^{BA}q_\mu q_\nu q_\rho q_\sigma(q+p)_\lambda$    &   $-\frac{i}{128 \pi ^2
   }\left(\frac{ \cN^\lambda_{AB} p_\lambda \cD^{\mu\ \nu}_{\ \mu\ \nu}{}^{BA}}{4}-\cN^\rho_{AB} p_\mu \cD^{\mu\ \nu}_{\ \rho\ \nu}{}^{BA}\right)$   \\ \hline
    $\cV\cV$     &   $-\frac12 \cV^{\mu\nu}_{AB} \cV^{\rho\sigma}{}^{BA} q_\mu q_\nu  (q+p)_\rho(q+p)_\sigma$     &   $\frac{1}{384 \pi ^2
   }\left( \cV^{\mu\nu}_{AB}
   \cV_{\mu\nu}{}^{BA}+\frac{ \cV^\mu_{\ \mu}{}_{AB} \cV^\nu_{\ \nu}{}^{BA}}{2}\right)$   \\ \hline
    $\cV \cC$     &    $-\frac i2 \cV^{\sigma\lambda}_{AB} \cC^{\mu\nu\rho}{}^{BA} q_\mu q_\nu q_\rho (q+p)_\sigma (q+p)_\lambda$    &  \begin{tabular}[c]{@{}l@{}}$-\frac{i}{256 \pi ^2
   }\Big(\frac{ p_\mu \cV^\nu_{\ \nu}{}_{AB} \cC^{\mu\rho}_{\ \ \rho}{}^{BA}}{2}+ p_\mu
   \cV_{\nu\rho}{}_{AB} \cC^{\mu\nu\rho}{}^{BA}$ \\ $ -p_\mu
   \cV^{\mu\nu}_{AB} \cC^{\ \rho}_{\nu\ \rho}{}^{BA}\Big)$\end{tabular}        \\ \hline
    $\cV\cD$     &    $\frac12 \cV^{\lambda\delta}_{AB} \cD^{\mu\nu\rho\sigma}{}^{BA}q_\mu q_\nu q_\rho q_\sigma(q+p)_\lambda(q+p)_\delta$    &   \begin{tabular}[c]{@{}l@{}}$\frac{1}{160 \pi ^2
   }\Big(-\frac{3  \cV^{\mu\nu}_{AB} p_\mu p_\nu  \cD^{\rho\ \sigma}_{\ \rho\ \sigma}{}^{BA}}{16}+p_\mu p_\nu \cV^\mu_{\ \rho}{}_{AB} \cD^{\nu\rho\sigma}_{\ \ \ \sigma}{}^{BA}-$\\ $\frac{3 p_\mu p_\nu \cV^\rho_{\ \rho}{}_{AB} \cD^{\mu\nu\sigma}_{\ \ \ \sigma}{}^{BA}}{8 }-\frac{3 p_\mu p_\nu \cV_{\rho\sigma}{}_{AB} \cD^{\mu\nu\rho\sigma}{}^{BA}}{4}+$\\ $\frac{p^2 \cV^\mu_{\ \mu}{}_{AB} \cD^{\nu\ \rho}_{\ \nu\ \rho}{}^{BA}}{16 }+\frac{p^2 \cV_{\mu\nu}{}_{AB}\cD^{\mu\nu\rho}_{\ \ \ \rho}{}^{BA}}{4}\Big)$ \end{tabular}       \\ \hline
    $\cC\cC$     & \begin{tabular}[c]{@{}l@{}}   $-\frac12 \cC^{\mu\nu\rho}_{AB} \cC^{\sigma\lambda\delta}{}^{BA} q_\mu q_\nu q_\rho \times$\\$(q+p)_\sigma(q+p)_\lambda(q+p)_\delta$ \end{tabular}       &  \begin{tabular}[c]{@{}l@{}}$\frac{1}{640 \pi ^2
   }\Big(-3  p_\mu p_\nu \cC^{\mu\rho\sigma}_{AB}\cC^\nu_{\ \rho\sigma}{}^{BA}-\frac{3
    p_\mu p_\nu \cC^{\mu\rho}_{\ \ \rho}{}_{AB}\cC^{\nu\sigma}_{\ \ \sigma}{}^{BA}}{2}-$\\ $  \frac{
    p^2 \cC^{\mu\nu\rho}_{AB}\cC_{\mu\nu\rho}{}^{BA}}{2}-\frac{3
    p^2 \cC^{\mu\ \nu}_{\ \mu}{}_{AB}\cC^{\ \rho}_{\nu\ \rho}{}^{BA}}{4}+  \frac{9
    p_\mu p_\nu \cC^{\mu\nu\sigma}_{AB}\cC^{\ \rho}_{\sigma\ \rho}{}^{BA}}{2}\Big)$\end{tabular}      \\ 
    \hline
    $\cC\cD$     &  \begin{tabular}[c]{@{}l@{}}  $-\frac i2  \cC^{\lambda\delta\alpha}_{AB} \cD^{\mu\nu\rho\sigma}{}^{BA}q_\mu q_\nu q_\rho q_\sigma\times$\\$(q+p)_\lambda(q+p)_\delta(q+p)_\alpha$ \end{tabular}       & \begin{tabular}[c]{@{}l@{}}$\frac{i}{320 \pi ^2
   }\Big(\frac{3 \cC^{\rho\sigma\delta}_{AB} \cD^{\mu\nu}_{\ \ \sigma\delta}{}^{BA}p_\mu p_\nu p_\rho}{2 }+\frac{3  \cC^{\rho\sigma}_{\ \ \sigma}{}_{AB} \cD^{\mu\nu\delta}_{\ \ \ \delta}{}^{BA}p_\mu p_\nu p_\rho}{4}-$ \\ $\frac{3
   \cC^{\mu\nu}_{\ \ \sigma}{}_{AB} \cD^{\rho\sigma\delta}_{\ \ \ \delta}{}^{BA}p_\mu p_\nu p_\rho}{2}+\frac{ \cC^{\mu\nu\rho}_{AB} \cD^{\sigma\ \delta}_{\ \sigma \ \delta}{}^{BA}p_\mu p_\nu p_\rho}{4}-$\\$ \cC^{\sigma}_{\ \sigma\delta}{}_{AB} \cD^{\mu\nu \rho\delta}{}^{BA}p_\mu p_\nu p_\rho- \frac{3 p^2 p_\mu
    \cC^{\mu\nu\rho}_{AB} \cD^{\ \ \delta}_{\nu\rho \ \delta}{}^{BA}}{4}-$\\$\frac{3 p^2 p_\mu
    \cC^{\mu\nu}_{\ \ \nu}{}_{AB} \cD^{\rho \ \sigma}_{\ \rho \ \sigma}{}^{BA}}{16}+\frac{p^2 p_\mu
    \cC^{\nu\rho\sigma}_{AB} \cD^{\mu}_{\ \nu\rho\sigma}{}^{BA}}{2}+$\\$\frac{3 p^2 p_\mu
    \cC^{\nu\ \sigma}_{\ \nu}{}_{AB} \cD^{\mu\ \delta}_{\ \sigma \ \delta}{}^{BA}}{4}\Big)$\end{tabular}        \\ \hline
    $\cD\cD$     &   \begin{tabular}[c]{@{}l@{}}  $-\frac12   \cD^{\mu\nu\rho\sigma}_{AB}\cD^{\lambda\delta\alpha\beta}{}^{BA}q_\mu q_\nu q_\rho q_\sigma\times$\\$(q+p)_\lambda(q+p)_\delta(q+p)_\alpha(q+p)_\beta$    \end{tabular}   &     \begin{tabular}[c]{@{}l@{}} 
    $-\frac{1}{140 \pi ^2
   }\Big(-\frac{9\cD^{\mu\nu}_{\ \ \alpha\beta}{}_{AB}\cD^{\rho\sigma\alpha\beta}{}^{BA}p_\mu p_\nu p_\rho p_\sigma}{16}-$\\$\frac{9 \cD^{\mu\nu\lambda}_{\ \ \ \lambda}{}_{AB}\cD^{\rho\sigma\delta}_{\ \ \ \delta}{}^{BA}p_\mu p_\nu p_\rho p_\sigma}{32}+ \cD^{\mu\nu\rho\delta}_{AB}\cD^{\sigma\ \lambda}_{\ \delta\ \lambda}{}^{BA}p_\mu p_\nu p_\rho p_\sigma-$\\$\frac{10 \cD^{\mu\nu\rho\sigma}_{AB}\cD^{\lambda\ \delta}_{\ \lambda \ \delta}{}^{BA}p_\mu p_\nu p_\rho p_\sigma}{64}-\frac{ 3 p^2 p_\mu p_\nu \cD^{\mu\rho\sigma\lambda}_{AB}   \cD^{\nu}_{\ \rho\sigma\lambda}{}^{BA}}{8}-$\\$\frac{9 p^2 p_\mu p_\nu \cD^{\mu\rho\ \lambda}_{\ \ \rho}{}_{AB}   \cD^{\nu\ \sigma}_{\ \lambda\ \sigma}{}^{BA}}{16}+\frac{3 p^2 p_\mu p_\nu \cD^{\mu\nu\rho\sigma}_{AB}   \cD^{\ \ \lambda}_{\rho\sigma\ \lambda}{}^{BA}}{4 }+$\\$\frac{3 p^2 p_\mu p_\nu \cD^{\mu\nu\rho}_{\ \ \ \rho}{}_{AB}  \cD^{\sigma\ \lambda}_{\ \sigma\ \lambda}{}^{BA}}{16}-\frac{3 p^4 \cD^{\mu\nu\rho\sigma}_{AB}\cD_{\mu\nu\rho\sigma}{}^{BA}}{128}-$\\$\frac{9 p^4 \cD^{\mu\ \rho\sigma}_{\ \mu}{}_{AB}\cD^{\ \ \nu}_{\rho\sigma\ \nu}{}^{BA}}{128}-\frac{9 p^4 \cD^{\mu\ \nu}_{\ \mu\ \nu}{}_{AB}\cD^{\rho\ \sigma}_{\ \rho\ \sigma}{}^{BA}}{1024 }\Big)$ 
   \end{tabular} \\
   \hline
\end{tabular}
\caption{Bubble loops for the graviton field.}
\label{tab}
\end{table}

Now let us consider the graviton field. Given a quadratic operator acting on a rank 2 symmetric tensor field
\be
\cO=\boxdot^2\mathbb{I}
+\cD^{\mu\nu\rho\sigma}\partial_\mu\partial_\nu\partial_\rho\partial_\sigma
+\cC^{\mu\nu\rho}\partial_\mu\partial_\nu\partial_\rho
+\cV^{\mu\nu}\partial_\mu\partial_\nu
+\cN^\mu\partial_\mu+\cU\ ,
\label{operatorff}
\ee
the contribution from bubble diagrams to the one-loop effective action reads
\be
\cU\cU+\cN\cN+\cV\cV+\cC\cC+\cD\cD+2(\cU\cV+\cU\cN+\cU\cC+\cU\cD+\cN\cV+\cN\cC+\cN\cD+\cV\cC+\cV\cD+\cC\cD),\label{sum}
\ee
where each term $\mathcal{A}\mathcal{B}$ in the sum corresponds to a bubble Feynman diagram composed by vertices  $\mathcal{A}$ and $\mathcal{B}$ from (\ref{operatorff}).
We introduce a generalized index notation for symmetric rank-2 tensors, $h_A:=h_{\mu\nu}$ and $\hat{1}=\delta_{AB}$. 
The operator $\cO$ and all its coefficients are matrices in the space of symmetric
tensors, so they carry hidden indices $A,B$. By convention such indices always
come after the ones contracted with derivatives, e.g. $\cV^{\mu\nu AB}$.
Terms proportional to $\log{p^2}$ from each of these diagrams are reported in table \ref{tab}.

Taking as $\cO$ the operator $\Box^2$ expanded at first order in $f$ with respect to the perturbed metric, 
$g_{\mu\nu}=\eta_{\mu\nu}+f_{\mu\nu}$, one finds
\be
-\frac{11}{48\pi^2}\br^{\mu\nu}\br_{\mu\nu}+\frac{7}{96\pi^2}\br^2\ .
\label{kindiv}
\ee
Since $\tr\log(\Box^2)=2\tr\log\Box$, we would expect that the result
must be equal to
\be
2\frac{1}{16\pi^2}b_4(\Box)=-\frac{R^{\mu\nu}R_{\mu\nu}}{144\pi^2} - \frac{ R^{\mu\nu\rho\sigma}R_{\mu\nu\rho\sigma} }{18\pi^2} + \frac{5R^2}{288\pi^2}
\label{hkbox}
\ee
(where $b_4$ is the heat kernel coefficient).
This expression is indeed equal to (\ref{kindiv}) if one takes $R^{\mu\nu\rho\sigma}R_{\mu\nu\rho\sigma}=4 R^{\mu\nu}R_{\mu\nu}-R^2$, which is equivalent to adding a total derivative term.

Taking instead $\cO$ to be the full operator given by Eq.~(\ref{operator}), expanded around flat space,
leads to 
\be
\frac{1617\lambda-20\xi}{1440\pi^2\lambda}\br^{\mu\nu}\br_{\mu\nu}
-\frac{37800\lambda^3+540\lambda^2\xi+3219\lambda\xi^2-40\xi^3}{8640\pi^2\lambda\xi^2}\br^2\ ,
\ee
that lead to our final result for the beta functions, Eqs.~(\ref{bdmpl}) and~(\ref{bdmpx}).

\end{document}